# Mixed-level identification of fault redundancy in microprocessors

Adeboye Stephen Oyeniran, Raimund Ubar, Maksim Jenihhin, Cemil Cem Gürsoy, Jaan Raik
Tallinn University of Technology, Estonia
adeboye.oyeniran@ttu.ee, raiub@pld.ttu.ee, maksim@pld.ttu.ee, cem@ati.ttu.ee, jaan@pld.ttu.ee

*Abstract*— **A new high-level implementation independent functional fault model for control faults in microprocessors is introduced. The fault model is based on the instruction set, and is specified as a set of data constraints to be satisfied by test data generation. We show that the high-level test, which satisfies these data constraints, will be sufficient to guarantee the detection of all non-redundant low level faults. The paper proposes a simple and fast simulation based method of generating test data, which satisfy the constraints prescribed by the proposed fault model, and a method of evaluating the high-level control fault coverage for the proposed fault model and for the given test. A method is presented for identification of the high-level redundant faults, and it is shown that a test, which provides 100% coverage of non-redundant high-level faults, will also guarantee 100% non-redundant SAF coverage, whereas all gate-level SAF not covered by the test are identified as redundant. Experimental results of test generation for the execution part of a microprocessor support the results presented in the paper.**

## I. INTRODUCTION

Technology scaling in today's deep-submicron processes produce new failure mechanisms in electronic devices, which has forced researchers to develop more advanced fault models compared to the traditional *stuck-at fault* (SAF) model [1], and to investigate the possibilities of reasoning the faulty behavior of systems without using any particular fault models [2, 3].

Fault models for digital circuits have been developed for different types of failure mechanisms like signal line *bridges* [4], transistor *stuck-opens* [5] or failures due to increasing circuit *delays* [6]. Another trend has emerged to develop general fault modeling mechanisms and corresponding test tools that can effectively analyze arbitrary fault types. The oldest example is the *D-calculus* [7]. A generalization of this approach has been found in the *input pattern fault model* [8], and in the *pattern fault model* [9], which can represent any arbitrary change in the logic function of a circuit block, where a block is defined to be any combinational sub-circuit described at any level of the design hierarchy.

A similar pattern related fault modeling approach called *functional fault model* was proposed earlier in [10] for the module level fault diagnosis in combinational circuits. The *functional* (or *pattern*) fault model allows an arbitrary set of signal lines to be grouped into activation conditions for a single fault site, allowing a variety of *physical defect* types to be modeled. Based on the *functional fault model* a deterministic *defect-oriented* test pattern generator DOT was developed in [11] which allowed proof of the logic redundancy of not detected physical defects.

In [12], a similar model called *conditional faults* was proposed for test generation purposes, and in [13] for diagnosis purposes. A conditional fault allows additional signal line objectives to be combined with the detection requirements of a particular fault. For complete exercising blocks in combinational circuits on the gate level, a similar pattern oriented *gate-exhaustive fault model* was proposed in [14], which was extended to target bigger regions (collections of gates) by *region-exhaustive fault* model in [15].

The described functional, conditional and pattern fault models offer high flexibility in defect modeling beyond single SAF model. Further advancements of the low-level fault modeling have been achieved by introducing the fault *tuple fault* model [16], *realistic sequential cell fault* model [17], or *cell-internal defect* model [18], where the last two cases provide general capability to handle sequential misbehavior of circuits.

The conditional SAF model (and other listed models) [8-18] support *hierarchical test approach*, where the test pattern (or sequence), which activates a low-level fault (e.g. physical defect) at the lower level can be considered as the high-level *condition* (or *constraint*) for the functional fault defined at the higher level.

To increase the speed of test generation and fault coverage evaluation, high-level (functional or behavioral) fault models have been developed. Such a model can be considered as "good", if the tests generated using this model provide a high coverage of SAF or physical defects.

In the design hierarchy, higher-level descriptions have fewer implementation details, but more explicit functional information than lower level descriptions. High-level fault models depend on which level the tests are generated.

Usually, the methods of high-level test generation are divided into structural RTL based methods [19-20], or behavioral test generation methods [21-22]. A high-level fault model can be explicit or implicit [23-24]. An explicit model identifies each fault individually, and every fault in this model will be a target for test generation. Implicit models are based on the assumption that all gate-level faults may not be represented at the RT level, and this motivated to develop dedicated RTL fault models with dependence on implementation details.

High-level fault models are used widely in the field of Software-Based Self-Test [25-30]. These approaches can be divided into two major groups - structural and functional. Structural approaches, such as [25-26], are based on test generation using information from lower level of design (gate- or RTL-level description) of processor under test. Functional, in its turn, is using instruction set architecture (ISA) information of the processor under test [27-30].

The main and general problem of high-level faults is the difficulty of proving that the model covers all low-level detectable (non-redundant) faults. In existence of such a high-level proof, it would be possible to identify the redundancy of gate-level faults exclusively by only gate-level fault simulation, which has cheaper cost than low-level fault redundancy proof by conventional gate-level ATPG-s.

In this paper, we make such attempt for a restricted class of circuits with well-defined functionality. Particularly, we target ALU control circuits. We propose a high-level data constraint based functional control fault model, and we prove that the test producing 100% high-level fault coverage will also guarantee 100% low-level detectable SAF coverage, and that all not detected SAF, identified by low-level fault simulation, are redundant.

The rest of the paper is organized as follows. In section 2, we present a novel control fault model for microprocessors, and in Section 3 we investigate the problem of mapping these high-level faults to low-level. Section 4 discusses high-level fault coverage measurement. In Section 5, we investigate the problem of high-level fault redundancies, and in Section 6 low-level fault redundancies. Section 7 presents experimental data, and Section 8 concludes the paper.

II. HIGH-LEVEL CONTROL FAULT MODEL FOR PROCESSORS

In this paper, we focus on testing of the ALU control, as a part of all control circuits in microprocessor cores.

Assume, the ALU executes $n$ different functions $y = f_i(D_i)$ by a set $F = \{f_i\}$ of instructions, where $D_i$ is the set of data operands for $f_i$, the length of the data word is $m$, and ALU is controlled by $p$ control signals. Consider a general ALU model partitioned into the control and data parts as shown in Fig.1. The control part consists of the multiplexer MUX and $p$ control lines as control inputs to MUX. The $n$ AND blocks in MUX have each $p$ control and a single m-bit data input, whereas the OR block in MUX has $n$ data word inputs from the outputs of AND blocks. Each AND block consists of $m$ AND gates with $p$ control inputs, and a single bit data input.

Let us classify two types of high-level functional fault models for ALU: *control faults* (the faults related to the control part of ALU), and *data faults* (the faults related to the data part of ALU). In the following, we will consider the control fault testing.

**Definition 2.** Introduce for the function (instruction) $f_i \in F$, the following *high-level control fault model* $CFM(f_i) = \{Ex(f_i), C(y_i, F)\}$, where $C(y_i, F)$ is a set of the constraints to be satisfied for each bit $k$ of $y_i$:

$$\forall k: (y_{i/k} \neq 0), \quad (1)$$

$$\forall f_j \in F, j \neq i : \{\forall k: (y_{i/k} < y_{j/k})\} \quad (2)$$

Depending on the technology, implemented in the microprocessor, the constant 0 in formula (1) can be changed into 1, and instead of the relation " < " in formula (2), there can be " > ".

**Definition 1.** Let us introduce *control fault universe* as a set of any multiple SAF and bridging faults on the control lines of the control part (shown as control fault locations in Fig.1).

Introduce the following notations: $Ex(f_i)$ – execution of the instruction $f_i$, $y_i$ – the data word considered as the result of the introduction $f_i$ at the data operands $D_i$.

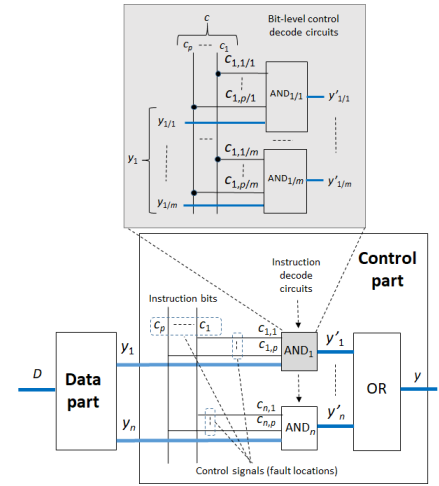

*Fig.1.* Generic DNF based control structure of ALU

The proposed fault model can be regarded as a generalization of the *conditional SAF model* (or similar ones considered in [8-18]). In case of conditional SAF, we are testing SAF on the gate-level lines at some constrained signals on other lines, whereas in case of the high-level fault model of Definition 2, we are testing the instructions of microprocessors at a set of constraints for data (operands).

Let us compare the complexities of the proposed high-level control fault model and the traditional SAF model of the control part architecture in Fig.1. The complexity of SAF model can be represented by the size of the model, i.e. by the number of control lines in the circuit multiplied by two for both SAF types: $C(SAF) = 2nmp$. On the other hand, the complexity of the proposed high-level control fault model (CFM) can be represented by the number of data constraints to be satisfied, that is $C(CFM) = n(n-1)mp$.

The time costs *TC* of test generation for both cases of the fault model can be estimated roughly by multiplying the size of the model with average test generation time *t* per fault as

$$TC(SAF) = 2nmp * t_{SAF} \quad (3)$$
$$TC(CFM) = n(n-1)mp * t_{FFM} \quad (4)$$

Despite that the size *C*(SAF) is linear with the circuit size, the gate-level test generation time with ATPGs is not scalable. On the other hand, despite the quadratric size of the high-level control fault model *C*(CFM), the test generation time for solving the data constraints (2) is linear, and very efficiently executable by random search (see experimental data).

Note, to simplify the model proposed in Definition 2 with the goal to reduce its size, the set of instructions *F* can be partitioned into subsets of *F*, similarly as proposed in [29], and for each subset, dedicated high-level fault model according to Definition 2 can be derived.

The practical reason for such partitioning of *F* may result from the instruction coding scheme. For example, if different fields of the instruction format have separate decoding circuit, then for each field, a separate set of instructions *F* for the proposed in Definition 2 fault model can be assigned.

III. HIGH-LEVEL CONTROL FAULT MAPPING TO GATE-LEVEL FAULTS

Let us consider in the following, how the gate-level fault redundancies in the control part of ALU can be identified by mixed level fault reasoning. To reduce the complexity of the problem, we propose, instead of exploiting slow conventional gate-level ATPGs for SAF redundancy proof, to use the combination of faster high-level test generation, and faster than ATPG low level fault simulation to achieve the same result – identification of the redundant low-level faults.

Introduce the following notations of the input information for solving the problem.

**Definition 3.** The set *CFM* = {*CFM*($f_i$)} for all $f_i \in F$ represents the full high-level control fault model for the given set of functions *F* = {$f_i$} of the microprocessor.

**Definition 4.** Let $D^*_i$ – be the set of data operands which satisfy the constraints of the fault model *CFM*($f_i$), $T^*_i$ – the test, which uses the data operands $D^*_i$, and $T^* = \{T^*_i\}$ – the full test, generated for the control fault model *CFM*.

**Theorem 1.** The test $T^* = \{T^*_i\}$, which covers all non-redundant high-level faults of the model *CFM* = {*CFM*($f_i$)}, covers also all gate-level testable SAF in the control part of the microprocessor, which controls the set of functions *F*.

Proof. Consider the generic ALU control part presented in Fig.1 and described as the following DNF:

$$y = c_{1,1}c_{1,2}\ldots c_{1,p}y_1 \vee c_{2,1}c_{2,2}\ldots c_{2,p}y_2 \vee \ldots \vee c_{n,1}c_{n,2}\ldots c_{n,p}y_n \quad (5)$$

In this DNF the variables $c_{i,j}$ for selecting the data results $y_i$, $i = 1,\ldots n$, represent the global control signals $c_j$, $j = 1,\ldots p$, being either inverted or not, and covering in general case exhaustively all the $2^p$ combinations. In DNF, according to Definition 4, and due to satisfied constraints (2) of the fault model in Definition 2, at least once the value of $y_i$ for each $i = 1,\ldots n$, will be $y_i = 1$. On the other hand, due to the exhaustiveness of all $2^p$ combinations of control signals, for each term of DNF with $y_i = 1$, there will be a combination of control signals $c_{i,1}c_{i,2}\ldots c_{i,p}$ consisting of a single 0, e.g. $c_{i,p} = 0$, with all others control signals $c_{i,r} = 1, r \neq p$. This is the case, where in the term $c_{i,1}c_{i,2}\ldots c_{i,p}y_i$, the SAF $c_{i,p} \equiv 1$ is activated. For propagating the fault $c_{i,p} \equiv 1$ to the output *y*, all other terms in DNF must have at least one 0 assigned to the variables of the term. This is guaranteed, because due to the constraints (2), which demands that in the term where all $c_{j,k} = 1$, the value of $y_j$ must be 0, and in other terms there must be at least one variable assigned by 0. Hence, all SAF faults of type $c_{i,p} \equiv 1$ in all variables $c_{i,p}$ can be tested by $T^*$.

The faults $c_{i,p} \equiv 0$ will be tested by patterns in $T^*$ where the constraint (1) is satisfied. ∎

**Corollary 1.** Any gate-level SAF in the control part related to *F* = {$f_i$}, not detectable by the test $T^* = \{T^*_i\}$ which covers all not redundant high-level control faults of the model *CFM* = {*CFM*($f_i$)}, is redundant.

Proof. In Theorem 1, exhaustiveness of using all the combinations of the local control signals $c_{i,1}c_{i,2}\ldots c_{i,p}$ was assumed. If not all combinations are used in the instruction set of the microprocessor, which is the typical practical case, then, not all patterns can be generated for activating all SAF of type $c_{i,p} \equiv 1$. Usually these cases are used for optimization of the gate-level structure of the control part of ALU. If however the optimization process has not removed all hardware redundancy, then as the result, the control part may consequently contain also redundant faults. These redundant faults can be identified by simple and fast gate-level fault simulation of the high-level generated test $T^*$. ∎

**Example 1.** Consider a simplified ALU unit with the set of three functions $f_1, f_2, f_3$, activated by a set of control signals $\bar{c}_2c_1, c_2\bar{c}_1, c_2c_1$ respectively. The ALU can be represented by the DNF:

$$y = \bar{c}_2c_1y_1 \vee c_2\bar{c}_1y_2 \vee c_2c_1y_3.$$

The test $T^* = \{T^*_1, T^*_2, T^*_3\}$ generated for the control part of ALU that satisfies the constraints (2) is depicted in Table 1.

*Table 1. Example of a high-level control test*

| $T^*_i$ | Test | | Fault table | | | Constraints satisfied |
|---|---|---|---|---|---|---|
| | $c_2 c_1$ | $y_1 y_2 y_3$ | $\bar{c}_2 c_1 y_1$ | $c_2 \bar{c}_1 y_2$ | $c_2 c_1 y_3$ | |
| 1 | 2 | | 3 | 4 | 5 | 6 |
| $T^*_1$ | 0 1 | 0 1 1 | 1 1 0 | 0 0 1 | 0 1 1 | $y_1 < y_2, y_1 < y_3$ |
| $T^*_2$ | 1 0 | 1 0 1 | 0 0 1 | 1 1 0 | 1 0 1 | $y_2 < y_1, y_2 < y_3$ |
| $T^*_3$ | 1 1 | 1 1 0 | 0 1 1 | 1 0 1 | 1 1 0 | $y_3 < y_1, y_3 < y_2$ |

The table contains the test patterns in column 2, the fault table in columns 3-5, and the constraints satisfied by generating data for the control test patterns in column 6. The detected gate-level faults in the fault table are highlighted by red colour: 0 means the value of a signal which activates the fault SAF/1. For example, in case of the fault $c_2 \equiv 1$ in column 5, the value of the output signal $y = y_1 = 0$ will change from 0 to $y = y_1 \vee y_3 = 1$. For detecting the faults SAF/0, more 3 test patterns are needed (not shown in the table). We see in the fault table that the faults $c_1 \equiv 1$ in column 3 and $c_2 \equiv 1$ in column 4 are not detected, because of the control code $c_2c_1 = 00$ is illegal (not usable in this ALU). According to

Corollary 1, these gate-level faults are redundant (in case if the control circuit is implemented as DNF). As the example shows, the redundancy of the gate-level faults can be derived by simple low-level SAF simulation. ∎

Note, Theorem 1 and Corollary 1 were formulated and the proofs were given, considering so far only the single SAF model. In fact, the power of the proposed high-level control fault model stretches far beyond the fault class of single SAF, as it will be shown in the following corollaries.

**Corollary 2.** The test $T^* = \{T^*_i\}$, covers all gate-level multiple SAF and bridging faults between control lines in the control part of the microprocessor, which controls the set of functions $F = \{f_i\}$.

Proof. From (2) it follows that for each function $F = \{f_i\}$, $\forall k: (y_{i/k} < y_{j/k})$ for all $j \neq i$ must hold. This means that not only SAF/1 in a single control signal of a single function $f_j \in F$, $j \neq i$, can be detected (by overwriting $y_{i/k} = 0$ with $y_{j/k} = 1$), where the control words for $f_i$ and $f_j$ differ in a single bit, rather such overwriting of signals $y_{i/k} = 0$ with 1 can happen, and hence, can be detected, due to multiple changes $0 \to 1$ for $f_j \in F$, $j \neq i$, leading to detecting multiple faults. This explanation can be derived also from reasoning of DNF (5).

On the other hand, from the constraints (1-2), and from the exhaustiveness of testing all the control functions function $f_j \in F$, $j \neq i$, it follows that non-redundant bridging faults between the control lines can also be detected by $T^*$. ∎

In case, when the target would be to detect only single SAF, then the fault model defined by the constraints (1) and (2) is over-dimensioned. For the case of full single SAF coverage, it would be sufficient to loosen the constraint (2) to

$$\forall f_j \in F, j \neq i, (HD(f_j, f_i) = 1) : \{\forall k: (y_{i/k} < y_{j/k})\} \quad (6)$$

where $HD(f_j, f_i) = 1$ is the constraint that the Hamming distance between the control codes for $f_j$ and $f_i$ is 1. This simplication is similar to the approach used in [29]

**Corollary 3.** The size of the proposed high-level control fault model applied only to the code-neighboring functions $f_j, f_i$ with $HD(f_j, f_i) = 1$, is equal to $C(CFM, HD=1) = nmp$.

Proof. The proof is straightforward, since for each $f_j \in F$, instead of $m(n-1)$, only $mp$ comparisons are needed. ∎

The size of the updated with (6) high-level control model is 2 times smaller than for the SAF model $C(SAF)$. Regarding the test cost, since $t_{CFM} \ll t_{SAF}$, we get

$$TC(CFM) = nmp \, t_{CFM} \ll TC(SAF) = 2nmp \, t_{SAF}. \quad (7)$$

## IV. HIGH-LEVEL FAULT COVERAGE MEASUREMENT

From above, it follows that the high-level control fault model CFM defined by the set of constraints (2) can be interpreted as the definition of the universe of high-level control faults. A direct impact of this interpretation is the possibility of evaluating the high-level fault coverage as the percentage of satisfied constraints (2) by the given test. The measuring of the coverage of constraints (1) is not needed, because they will be satisfied anyway as the byproduct of the data path test.

The size of the proposed high-level functional fault model results from the fault table for representing the coverage of satisfied constraints (2).

Let us introduce the high-level fault table as a matrix $D = ||D_{i,j}||$ with $n$ columns and $n$ rows, where $n$ – is the number of functions in $F$. Each entry $D_{i,j}$ in $D$ is a $m$-bit vector $D_{i,j} = (D_{i,j/1}, D_{i,j/2}, \ldots, D_{i,j/m})$, where $m$ is the number of bits in the data-word. $D_{i,j/k} = 1$, if the constraint $y_{i/k} < y_{j/k}$ for the bit $k$ is satisfied, and $D_{i,j/k} = 0$ if not.

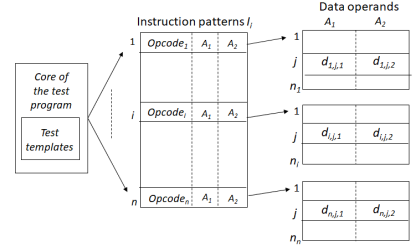

*Fig.2.* Architecture of the test program

Consider a simplified architecture of a test program for testing the control part of ALU as shown in Fig.2. The test $T^* = \{T^*_1, \ldots, T^*_n\}$ for ALU with $n$ functions of the set $F = \{f_1, \ldots, f_n\}$ consists of a core of the test program, array of test patterns (instructions) and array of test data operands. The core consists of a small set of test templates for initializing registers, executing test patterns and processing test results. The test patterns are instructions, and to each instruction, a set of data operands is assigned, to be exercised cyclically. Each test pattern with related operands forms a test $T^*_i \in T^*$. The task of the core is execution of the full test $T^*$.

For high-level fault simulation, there is no need to simulate the full test program illustrated in Fig.2. Instead of that, only the array of data operands should be processed according to the following procedure.

**Procedure 1.**
1) **for** $i = 1, \ldots, n$
2)    **for** all data operands $d_{i,j,1}, d_{i,j,2}, j = 1, \ldots, n_i$
3)       **for** all instructions $f_h$, $h = 1, \ldots, n$
4)          calculate the value $y_h$
5)          check the relation $y_i < y_h$, $h \neq i$
5)          update the vector $D_{i,h} \in D$
6) **end for**

For high-level test generation we developed a simulation based random search for test data to satisfy the constraints (2), where for constraint checking we used Procedure 1.

## V. IDENTIFICATION OF HIGH-LEVEL CONTROL FAULT REDUNDANCIES

Consider Table 2, which illustrates a fragment of the high-level fault coverage matrix $D$, for a test $T^*$ generated for the MiniMIPS processor [24]. In this fragment 8-bit data-words, and 5 functions OUI, ADD, SUB, SLT, AND of the MiniMIPS microprocessor are considered.

In Table 2, the 0s refer to the possible high-level redundancies of the control faults related to the constraints $y_{i/k} < y_{j/k}$, where $i$ and $j$ correspond to the rows and columns, respectively. All 0s in $D_{ij}$ refer to the high probability of redundancy of the full set of high-level faults for all bits, which means that the constraints $y_{i/k} < y_{j/k}$, for all $k$ cannot be satisfied. In most cases of ALU operations, it is very easy to identify this type of redundancy. For example, if $y_i = f_i(a, b)$ refers to AND operation and $y_j = f_j(a, b)$ refers to OR, it is straightforward that the constraint $y_i < y_j$, i.e. $(a \vee b) < (a \wedge b)$ cannot be satisfied.

**Table 2.** *Example of a High-Level Fault Table*

|            | $f_1$(OUI) | $f_2$(ADD) | $f_3$(SUB) | $f_4$(SLT) | $f_5$(AND) |
|------------|------------|------------|------------|------------|------------|
| $f_1$(OUI) |            | 111111     | 111111     | 111111     | **000000** |
| $f_2$(ADD) | 11111      |            | 11111**0** | 111111     | 111111     |
| $f_3$(SUB) | 11111      | 11111**0** |            | 111111     | 111111     |
| $f_4$(SLT) | 11111      | 111111     | 111111     |            | **000000** |
| $f_5$(AND) | 11111      | 111111     | 111111     | 111111     |            |

In Table 2, the 0s refer to the possible high-level redundancies of the control faults related to the constraints $y_{i/k} < y_{j/k}$, where $i$ and $j$ correspond to the rows and columns, respectively. All 0s in $D_{ij}$ refer to the high probability of redundancy of the full set of high-level faults for all bits, which means that the constraints $y_{i/k} < y_{j/k}$, for all $k$ cannot be satisfied. In most cases of ALU operations, it is very easy to identify this type of redundancy. For example, if $y_i = f_i(a, b)$ refers to AND operation and $y_j = f_j(a, b)$ refers to OR, it is straightforward that the constraint $y_i < y_j$, i.e. $(a \vee b) < (a \wedge b)$ cannot be satisfied.

In cases when there is an entry $D_{i,j/k} = 1$ in a single bit $k$ of the vector $D_{ij}$, or in only few bits of it, we can suggest for the proof a method called "*partial truth table method*". The idea of the method stands in showing the equivalence of partial truth tables (or to prove the impossibility of solving the related constraints) for the functions involved in the constraint relation, so that as few as possible responsible bits should be selected for the need of the proof.

In Table 3, examples are shown for 1-bit partial truth tables for the functions SUB, ADD, OR, AND, for bit $k$. The pairs 00, 01, 10, 11 represent the values of the data variables (as arguments) in bit $k$, and the 1-bit values in the columns show the results of the related operations for this $k$-th bit. For SUB and ADD, the equivalence of the behavior in the given bit is demonstrated, which contradicts to the constraint (2), and in the case of OR and AND, the missing of a solution for (2) is also shown for all possible input data combinations.

It is easy also to show for example, the equivalence of operations ASR and SHR for MiniMIPS for all bits, except the most significant bit MSB. Hence, for all bits except for MSB, the entry $d_{i,j/k} = 0$ refers to the redundant control fault.

In some cases, the partial truth table method will not work, because the results of operations may substantially depend on all bits of the word like for increment or decrement operations. When this happens, specific corner cases should be found for the proof of redundancy. For example, to prove the equivalence of increment and decrement operations in the least significant bit, the operand 1…110 should be used, where both instructions INC and DEC produce the same result "all 1s".

**Table 3.** *Examples of redundancy proofs with 1-bit truth tables*

| # | $y_{i/k} < y_{j/k}$ | $D_{ij}$ |     | 00 | 01 | 10 | 11 |
|---|---------------------|----------|-----|----|----|----|----|
| 1 | SUB < ADD           | 1…11**0**| SUB | 0  | 1  | 1  | 0  |
|   |                     |          | ADD | 0  | 1  | 1  | 0  |
| 2 | OR < ADD            | 1…11**0**| OR  | 0  | 1  | 1  | 1  |
|   |                     |          | ADD | 0  | 0  | 0  | 0  |

## VI. MIXED-LEVEL IDENTIFICATION OF FAULT REDUNDANCIES

Let us now draft the general procedure of the mixed-level identification of gate-level single SAF, where the test is generated at the high-level using the proposed high-level control fault model, and the redundancy of the low-level SAF is identified by low-level fault simulation of the test, generated at the high-level.

**Procedure 2.**

1) Generation of the high-level test $T^*$ for the given set of functions (instructions), with finding the data which satisfy the constraints (2) (see Section II).

2) Generation of the high-level fault coverage table $D$ by high-level fault simulation of the test $T^*$ (see Section IV). The steps 1 and 2 can be carried out jointly (see Section IV).

3) High-level fault redundancy identification. For all not covered high-level faults (all 0s in the fault table $D$), the redundancy of the high-level control faults is identified (see Section V).

4) If the high-level redundancy cannot be proven for some of high-level fault

s, the test $T^*$ must be extended to satisfy the constraints (2), and to achieve 100% high-level fault coverage. This is the prerequisite (Theorem 1) for the next step of redundant SAF identification.

5) Gate-level fault simulation of the test $T^*$. The not detected SAF are identified as redundant low level faults in the control circuit of ALU (Corollary 1).

## VII. EXPERIMENTAL RESULTS

We carried out experiments which consists of high-level test data generation for the control part of Execute stage of MiniMIPS processor [31], Fig.3. The test program generation included manual synthesis of test templates, high-level generation of test data (operands) to satisfy constraints (1-2), test program synthesis and high-level fault simulation. For high-level test generation and fault simulation we used home-made tools, whereas for gate-level operations we used commercial tool. Experimental results are shown in Table 4.

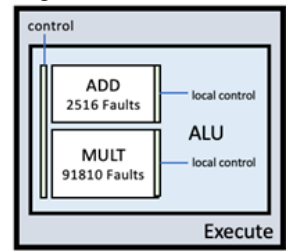

*Fig.3.* Simplified structure of the Execute stage of MiniMIPS

*Table 4. Experimental data*

| Approach | Experiments | | Faults | FC% | # Pat | ATPG time |
|---|---|---|---|---|---|---|
| Proposed high-level approach | High-level ATPG | | 756 | 100 | 196 | 47 s |
| | Gate-level simulation | ALU | 2516 | 99.92 | | |
| | | MULT | 91810 | 99.09 | | |
| Commercial gate-level ATPG | | ALU | 2516 | 99.96 | 169 | 1h 34 min |
| | | MULT | 91810 | 98.63 | | |

The experiments targeted ALU and MULT modules in the Execute stage of MiniMIPS. We generated a test with 100% coverage of high-level control faults. The operands generated according to (1-2), produced high gate-level SAF coverage for both, control and data parts of the Execute module.

The high-level test was simulated by commercial tool to grade the gate-level SAF coverage. To evaluate the efficiency of the high-level ATPG, we used for comparison also commercial gate-level ATPG. The time cost for high-level ATPG is about two orders of magnitude less than that of the commercial ATPG. The gate-level SAF coverage, achieved by the proposed ATPG for the whole module under test, is better than that achieved by the commercial tool.

The main goal of the experiments was to demonstrate the possibility of identification by high-level test generation the gate-level SAF redundancies. We demonstrated it on the basis of ALU test. The SAF coverage 99.92, achieved by 100% high-level fault coverage test, means that 2 faults remained in ALU not detected, and are qualified, according to Corollary 1, as redundant. Since by fault simulation of the test for ALU (without its local control part) we found 100% SAF coverage, we can conclude that the 2 faults belong to the ALU control part. On the other hand, since low-level ATPG found 1 undetected fault in the ALU joint data/control circuit, we can conclude that this redundant fault belongs to the ALU local control part, and the second redundant fault belongs to the ALU global control part (see Fig.3).

In the MULT block, fault coverage 99.09 refers to 835 not covered faults, which should be qualified according to Corollary 1 as redundant. By gate-level ATPG we found that from the 1256 not covered by ATPG faults, 865 were ATPG untestable, 105 were classified as redundant, and 286 remained not detected. From the latter it follows, that the 444 faults (the difference 1256 – 835), not covered by gate-level ATPG, however, were covered by the high-level ATPG. These faults should belong to the class of gate-level ATPG untestable faults.

## VIII. CONCLUSIONS

In this paper, we propose a novel high-level fault model which was experimented for test generation for ALU control parts in processors. The model consists of a set of data constraints to be satisfied by data operands that is to be used in the test. The constraints are derived from instruction set, in which case, no implementation details are needed. The test is able to detect all non-redundant single and multiple SAF, and bridging faults in the control circuit under test. Hence, the proposed method is more powerful than the traditional ATPGs, which target only single SAF. A metric and a method for high-level fault simulation with a method for identification of high-level fault redundancies were developed. We demonstrated the feasibility of the proposed method to identifying low level redundant SAF by combining high-level ATPG and low level SAF simulation. The test program generated explicitly for testing only the control part achieves as well a very high fault coverage for data part. This is due to the power of constraints (1-2) to be used for selecting data operands.

The future work will be to extend the proposed method for broader instruction sets of processors. Several optimization techniques are also possible.

**Acknowledgment:** The work has been supported by EU's H2020 project RESCUE, Estonian research grant IUT 19-1, and funded by Excellence Centre EXCITE in Estonia.